\documentclass[]{spie}  

\usepackage[top = 1 in, left = 1 in, right = 1 in, bottom=1.65 in]{geometry}
\usepackage[]{fancyhdr}
\usepackage{siunitx} 
\usepackage{amsmath,amsfonts,amssymb}
\usepackage{graphicx}
\usepackage[colorlinks=true, allcolors=blue]{hyperref}

\usepackage{amsmath,amssymb,amsfonts}
\usepackage{algorithmic}
\usepackage{graphicx}
\usepackage{textcomp}
\usepackage{xcolor}
\def\BibTeX{{\rm B\kern-.05em{\sc i\kern-.025em b}\kern-.08em
    T\kern-.1667em\lower.7ex\hbox{E}\kern-.125emX}}
\usepackage[utf8]{inputenc}
\usepackage{datetime}
\usepackage{upgreek}
\usepackage{dblfloatfix}    
\usepackage{subfigure}
\usepackage{booktabs}
\usepackage[font=small]{caption}
\usepackage{xcolor,soul,lipsum}
\usepackage{bm}
\usepackage{makecell} 
\usepackage{xr-hyper}
\usepackage{hyperref}
\hypersetup{
    colorlinks=true,
    linkcolor=blue,
    filecolor=magenta,      
    urlcolor=cyan,
}

\usepackage{threeparttable}
\makeatother
\usepackage[hyphenbreaks]{breakurl}
\usepackage[acronym]{glossaries}
\makeatletter

\usepackage[capitalize]{cleveref}
\crefname{section}{Sec.}{Secs.}
\Crefname{section}{Sec.}{Sects.}
\Crefname{table}{Table}{Tables}
\crefname{table}{Table}{Tabs.}

\newacronym{DVS}{DVS}{Dynamic Vision Sensor}
\newacronym{AER}{AER}{Address Event Representation}
\newacronym{EVS}{EVS}{Event-based Vision Sensor}
\newacronym{CMOS}{CMOS}{Combined Metal-Oxide Semi-conductor}
\newacronym{CTNU}{CTNU}{Contrast Threshold Non-Uniformity}
\newacronym{EPS}{EPS}{Events Per Second}
\newacronym{DN}{DN}{Digital Number}
\newacronym{COTS}{COTS}{Commercial Off The Shelf}
\newacronym{NCT}{NCT}{Nominal Contrast Threshold}
\newacronym{PDF}{PDF}{Probability Density Function}
\newacronym{ERC}{ERC}{Event Rate Controller}
\newacronym{FOV}{FOV}{Field of View}
\newacronym{ROI}{ROI}{Region of Interest}
\newacronym{AFOV}{AFOV}{Angular Field of View}
\newacronym{DR}{DR}{Dynamic Range}
\newacronym{CDF}{CDF}{Cumulative Distribution Function}
\newacronym{RP-TP}{RP-TP}{Reset Pulse-Test Pulse}
\newacronym{SW}{SW}{Square Wave}
\newacronym{LLCO}{LLCO}{Low Light Cutoff}

\title{Re-interpreting the Step-Response Probability Curve to Extract Fundamental Physical Parameters of Event-Based Vision Sensors}

\author[*a]{Brian J. McReynolds}
\author[a]{Rui Graca}
\author[b]{Lucas Kulesza}
\author[b]{Peter McMahon-Crabtree}
\affil[a]{Institute of Neuroinformatics, UZH/ETH Zurich, Winterthurerstrasse 190, 8057 Zürich, Switzerland}
\affil[b]{Air Force Research Laboratory, Space Vehicles Directorate, Albuquerque, NM, 87117 USA}

\authorinfo{Further author information: (Send correspondence to Brian McReynolds)\\Brian McReynolds: E-mail: brian.mcreynolds@afacademy.af.edu, Telephone: +41 77 947 6780\\  
}

\begin{document} 
\maketitle

\fancyfoot[C]{Approved for public release; distribution is unlimited. Public Affairs release approval \# AFRL-2024-1433.}

\begin{abstract}
Biologically inspired event-based vision sensors (EVS) are growing in popularity due to performance benefits including ultra-low power consumption, high dynamic range, data sparsity, and fast temporal response.  They efficiently encode dynamic information from a visual scene through pixels that respond autonomously and asynchronously when the per-pixel illumination level changes by a user-selectable contrast threshold ratio, $\theta$.  Due to their unique sensing paradigm and complex analog pixel circuitry, characterizing \gls{EVS} is non-trivial.  The step-response probability curve (S-curve) is a key measurement technique that has emerged as the standard for measuring $\theta$.  Though the general concept is straightforward, obtaining accurate results requires a thorough understanding of pixel circuitry and non-idealities to correctly obtain and interpret results. Furthermore, the precise measurement procedure has not been standardized across the field, and resulting parameter estimates depend strongly on methodology, measurement conditions, and biasing – which are not generally discussed. In this work, we detail the method for generating accurate S-curves by applying an appropriate stimulus and sensor configuration to decouple 2nd-order effects from the parameter being studied.  We use an \gls{EVS} pixel simulation to demonstrate how noise and other physical constraints can lead to error in the measurement, and develop two techniques that are robust enough to obtain accurate estimates. We then apply best practices derived from our simulation to generate S-curves for the latest generation Sony IMX636 and interpret the resulting family of curves to correct the apparent anomalous result of previous reports suggesting that $\theta$ changes with illumination. Further, we demonstrate that with correct interpretation, fundamental physical parameters such as dark current and RMS noise can be accurately inferred from a collection of S-curves, leading to more accurate parameterization for high-fidelity EVS simulations.                
\end{abstract}


\section{INTRODUCTION}
\label{sec:intro}  

\gls{EVS} employ a unique sensing paradigm inspired by biological vision. Their design captures dynamic visual information with high temporal resolution, low latency, wide dynamic range and limited power consumption, making them ideal for many diverse applications. These benefits have cemented \gls{EVS} as an important tool in computer vision and robotics \cite{Gallego2022-av}. In many such application areas, the goal is simply detecting the spatio-temporal location of an object or obstacle, and precise knowledge of \gls{EVS} operating parameters is of lesser importance. However, interest is growing in applying \gls{EVS} to scientific applications where more precise interpretation of \gls{EVS} output is needed.     
The most important \gls{EVS} performance parameter to characterize is contrast threshold, $\theta$. Fundamentally, \gls{EVS} operate on the principle of change detection--when a pixel senses a change its incident log-illumination level, it reports either an ON or OFF event, indicating respectively an increase or decrease in brightness. The change in log-illumination needed to trigger an event is the threshold. ON and OFF thresholds ($\theta_\text{on,off}$) are independent and able to be adjusted by the user to increase or decrease sensitivity, depending on sensing requirements. Various measurement techniques have been proposed to estimate $\theta$, with the most common being the event probability curve (S-curve) first introduced in \cite{Posch2011-vg}. The method consists of recording pixel response to periodic illumination pulses to a constant amplitude stimulus, then incrementally increasing the amplitude and calculating the response probability at each. This results in a curve that starts at zero and approaches 1 as the pulse amplitude reaches the threshold, and has emerged as the gold standard for reporting $\theta$ among new \gls{EVS} variants \cite{Finateu2020-kw,Niwa2023-io,Guo2023-ca}. Despite its popularity, the S-curve results in different estimates of $\theta$ at different brightness levels and bias settings, leading to uncertainty in the measurement. 

In this paper, we revisit the event probability S-curve with a holistic view of \gls{EVS} pixel operation and performance to examine the anomalous result that $\theta$ changes with illumination as reported in previous studies. We first use an \gls{EVS} pixel simulation to systematically study how measured S-curves relate to the pixel's true threshold(s). Starting with an ideal pixel model, we incrementally add non-ideal behaviors such as noise, finite bandwidth, and mismatch to show how measured S-curves are influenced. At each step, we describe how \gls{EVS} settings can be adjusted to reduce the impact of non-idealities, resulting in a prescribed measurement technique that is robust to noise and mismatch and a new interpretation of the S-curve that allows more accurate characterization of the contrast threshold. We demonstrate this method with the latest generation Sony IMX636 \gls{EVS}. Additionally, we use this new knowledge to re-interpret measurement results for the Sony IMX636 \gls{EVS}, and show that by properly interpreting reported S-curves, fundamental parameters such as dark current and noise can be inferred to facilitate more accurate simulations.


\section{EVS DESCRIPTION}
\label{sec: description}

\begin{figure} [ht]
   \begin{center}
   \includegraphics[width=\textwidth]{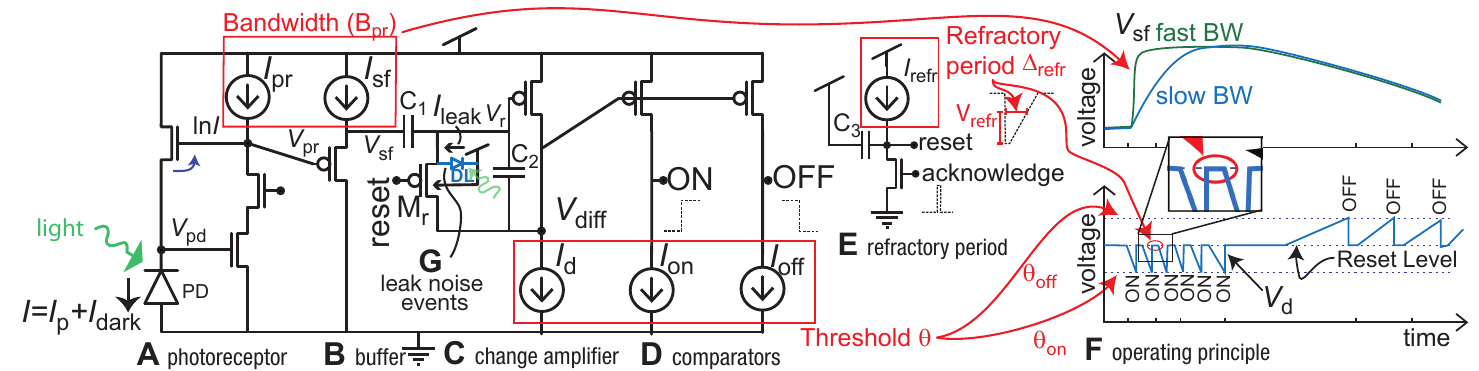}
   \end{center}
   \caption[example] 
   { \label{fig:block_diagram} 
Typical EVS pixel circuit schematic and description \cite{Graca2023-hk}. The active logarithmic photoreceptor front-end (\textbf{A-B}) drives a cap-feedback change amplifier (\textbf{C}) with output \textbf{$V_{diff}$}. When \textbf{$V_{diff}$} deviates by either an ON or OFF threshold, comparators (\textbf{D}) report an event, and after a finite refractory period (\textbf{E}) the change amplifier is reset. \textbf{F:} After each reset, the pixel again responds to signal changes around a new reference level. }
   \end{figure}

Each \gls{EVS} pixel operates autonomously and asynchronously, reporting `events' when the log-illumination level changes by a user defined threshold. To achieve this, each pixel contains a photodiode and analog circuitry to implement the change detection operation. \cref{fig:block_diagram} shows the standard pixel schematic introduced in \cite{Lichtsteiner2008-mm} and still used with minor modifications in commercial \gls{EVS} designs. The photoreceptor circuit logarithmically compresses the photocurrent to output $V_\text{pr}$. A source-follower buffer decouples the photoreceptor output from downstream circuit components and low-pass filters the signal to remove high frequency noise components. The signal is amplified by the well matched capacitor ratio $\frac{C_\text{1}}{C_\text{2}}$, and this signal is compared to the pixel's previous memorized brightness level. Independent ON and OFF comparators trigger if the signal change exceeds $\theta_\text{on}$ or $\theta_\text{off}$ respectively, and an `event' is recorded. After each event, the pixel is held in reset for a finite, adjustable period of time known as the refractory period. After the refractory period, the pixel memorizes a new brightness level which is used as a reference for reporting subsequent events. Each event is read out by peripheral circuitry, and \gls{EVS} output consists of a list of events, each containing a pixel address (x,y), timestamp (t), and a binary polarity output (p) indicating whether the event is an ON or OFF event.

\section{Step Response Probability Curve}
\begin{figure} [ht]
   \begin{center}
   \includegraphics[width = 400pt]{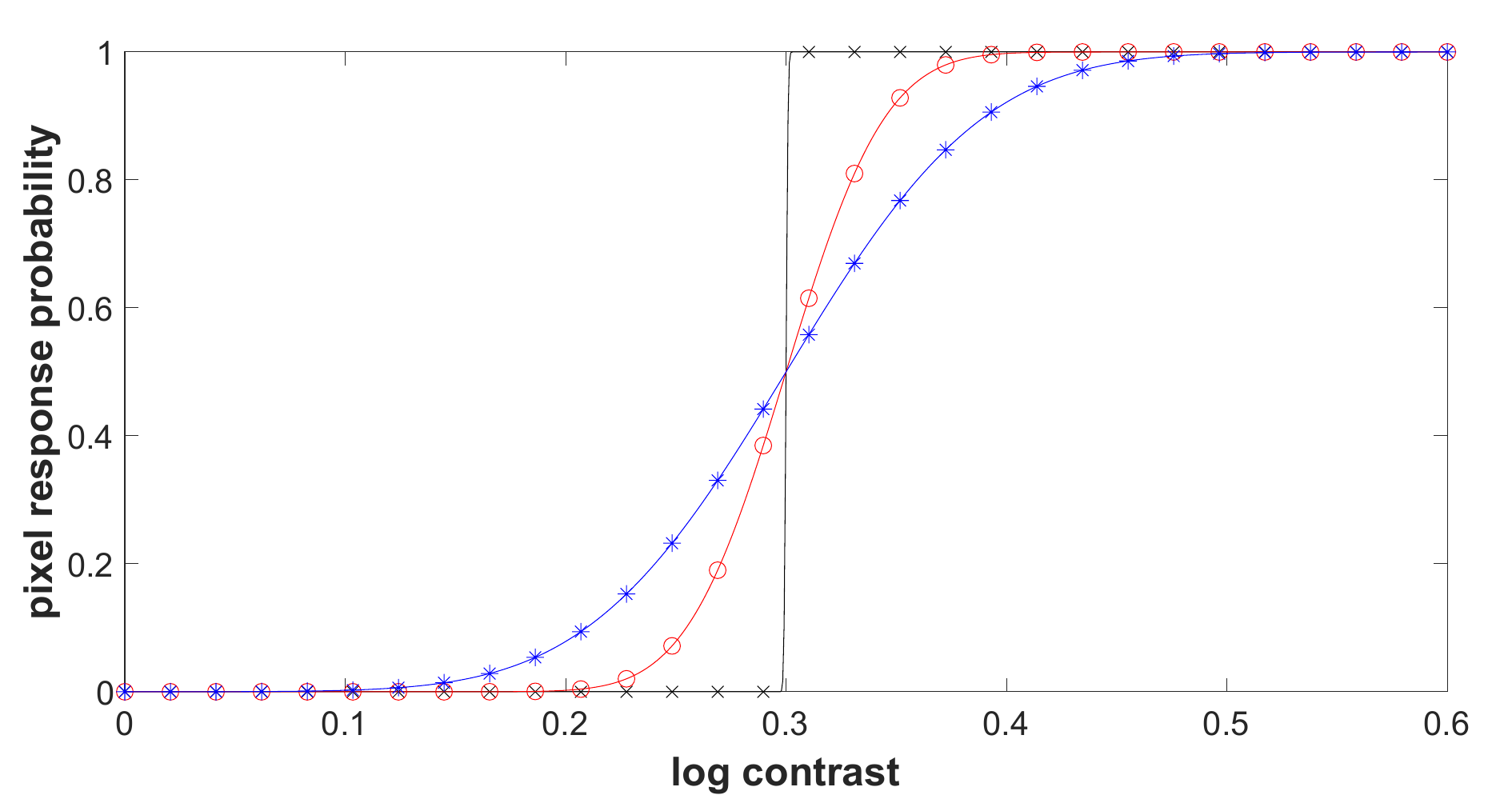}
   \end{center}
   \caption[example] 
   { \label{fig:example} Theoretical event probability curves for $\theta$ = 0.3.  In the ideal, noise-free case (black), the pixel response rises from 0 to 1 when the log contrast (amplitude) of the stimulus equals the contrast threshold. Due to noise, the response deviates from the ideal case, and more noise (blue curve) results in a larger deviation from ideal (shallowest slope). \gls{NCT} is defined as the 50\% event probability intercept, and is broadly accepted as the metric to characterize \gls{EVS} contrast threshold, $\theta$.  In this paper, we show that this interpretation does not result in an accurate measurement of $\theta$.
 }
   \end{figure}

The concept of using a step response probability curve (S-curve) to measure thresholds is simple. A group of pixels is exposed to a periodic stimulus of known amplitude, and event probability is calculated as the number of events recorded divided by the number of stimulus pulses. This procedure is repeated for several different stimulus amplitudes and event probability is plotted as a function of the log-contrast of the stimulus,
\begin{equation}
\label{eq:log_contrast}
\ \text{log contrast} = log\left(\frac{E_\text{v,max}}{E_\text{v,min}}  \right) \, ,
\end{equation} 
where $E_\text{v,max}$ and $E_\text{v,min}$ are the on-chip illuminance values of the visual stimulus. An example is shown in \cref{fig:example}. In the ideal case, this should result in a perfect step function--pulse amplitudes below the threshold result in zero response probability, and pulse amplitudes above the threshold result in 100\% response probability. Posch and Matolin were the first to propose this method and fit a Gaussian \gls{CDF} to the resulting curve \cite{Posch2011-vg}. Any deviation from an ideal step response is attributed to noise, and as a result the 50\% probability intercept is selected as the \gls{NCT}. This general measurement procedure and interpretation is broadly accepted throughout the \gls{EVS} community; however, McReynolds et al. studied this method using node voltages from a single \gls{EVS} test pixel and proposed that the true contrast threshold is near the 100\% response probability intercept \cite{McReynolds2022-fq}. The nature of the stimulus is another important consideration. \cite{Posch2011-vg} describes a more complicated stimulus consisting of a reset pulse preceeding each test pulse in order to ensure the pixel reference resets to a desired baseline level before each test pulse, but specific details of suggested reset pulse amplitude and duration are not discussed. It is logical that this procedure should produce more consistent/accurate results, but the amplitude of the reset pulse and duration of both reset and test pulses must be carefully selected. In our analysis, we will examine pixel response both with and without a reset pulse.          

\section{Experimental Method}

To assess the accuracy of the step response probability curve, we devised a test based on a pixel simulation similar to v2e \cite{Hu2021-df}. The key benefits of using a simulation are: 1) ground truth is known with 100\% accuracy, and 2) multiple stimulus waveforms and bias parameter combinations can be rapidly tested to develop best practices and assess robustness of each. Our simulation takes an input signal, compresses it logarithmically, adds Gaussian white noise, low pass filters the response, and implements event generation logic by comparing each time sample to the pixel's memorized reference level. If the sample deviates from the reference level by either the ON or OFF threshold, an event is recorded. We also include a refractory period, which is necessary to capture realistic pixel behavior. After each event, the simulation skips the appropriate number of samples corresponding to the length of the refractory period before resetting its reference. After reset, the event generation logic recommences. The low-pass filter is implemented by Euler-stepping the discretized differential equation corresponding to a first-order low-pass filter,   
\begin{equation}
\label{eq:iir_lowpass}
\ V_\text{out}[n] =  V_\text{out}[n-1] + \frac{\Delta t}{\tau}\left(V_\text{in}[n] - V_\text{out}[n-1]\right)\, .
\end{equation} 
In this case, $\tau = \frac{1}{2\pi f_\text{3dB}}$, where $f_\text{3dB}$ is the low-pass corner frequency. For the study, we used a time resolution ($\Delta t$) of 10$\mu$s, which is still many times faster than the highest low-pass bandwidth tested (2kHz). Appropriate values of noise power were determined by recording thirty separate 100 second simulations with a DC input signal. Noise power was then fine-tuned to obtain realistic target background activity rates, measured as events per pixel per second (typically reported in Hz). We selected $f_\text{3dB}$ values of 2kHz, 200Hz, and 50Hz, within the range of potential source-follower bandwidth settings, and for each, found noise power levels resulting in average background activity rates 0.5 Hz and 0.02Hz.       

\section{Results}

\begin{figure} [ht]
   \begin{center}
   \includegraphics[width = \textwidth]{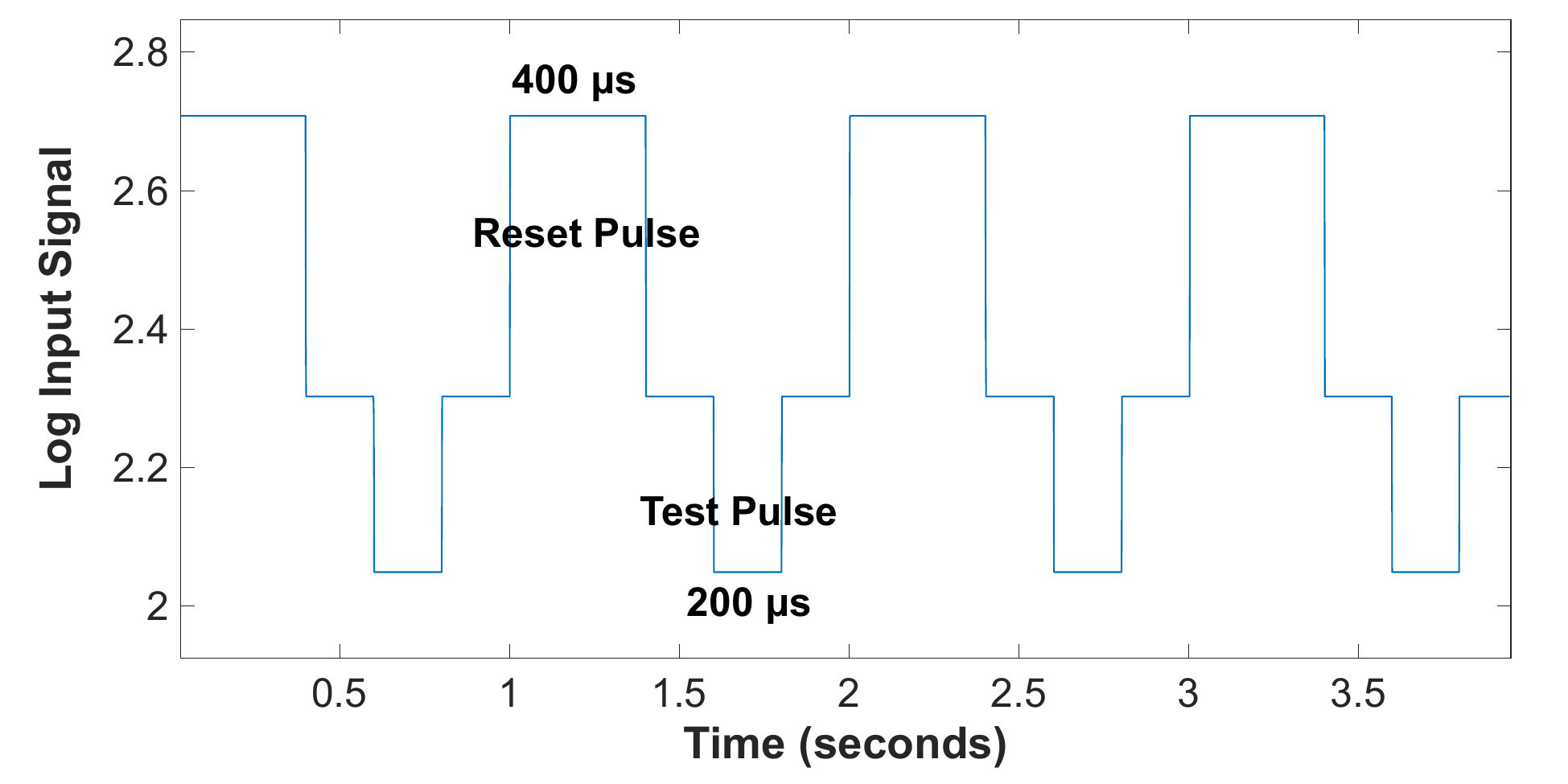}
   \end{center}
   \caption[example] 
   { \label{fig:RP_example} Sample reset pulse - test pulse waveform for testing OFF threshold. A 400 ms duration, 50\% contrast reset pulse precedes each test pulse. OFF events that occur during each test pulse are summed to determine event probability at a particular contrast level (test-pulse amplitude).
 }
   \end{figure}

In this section, we examine the S-curve response to two different stimulus patterns: 1) a simple periodic square wave with 50\% duty cycle, and 2) a \gls{RP-TP} sequence, which is a series of alternating rectangular reset pulses and test pulses. For the periodic square wave stimulus, we selected a 5 Hz modulation frequency. We tested various combinations of reset pulse amplitudes and frequencies, and eventually selected a 50\% contrast reset pulse amplitude, $\approx$ 55\% higher than the contrast threshold in log-signal units, with a 400 ms pulse duration, and 200 ms pulse duration for each test pulse. A sample reset-test pulse train is shown in \cref{fig:RP_example}. In the case of the reset pulse method, only events that occur within the temporal window of each test pulse are used to construct the S-curve. For each S-curve measurement, we apply 30 different contrast amplitudes ranging from 0.01 to 0.7 (1-70\%) linear contrast. In the remainder of this chapter, we systematically analyze both methods and assess how accurately they characterize contrast threshold. We start with the ideal case, then incrementally add in noise, finite bandwidth, and mismatch to demonstrate how the simulated S-curve measurement evolves. Where appropriate, we demonstrate biasing techniques that can be used to manage non-ideal behaviors and restore the integrity of the measurement. Finally, we compare results for the two stimuli, prescribe a complete methodology for obtaining accurate threshold measurements, and establish limits on acceptable noise levels over which accurate results can be attained. For all measurements, we set the ON and OFF thresholds, $\theta$, at 0.3 unless otherwise noted   

\subsection{Ideal Case}

\begin{figure} [ht]
   \begin{center}
   \includegraphics[width = \textwidth]{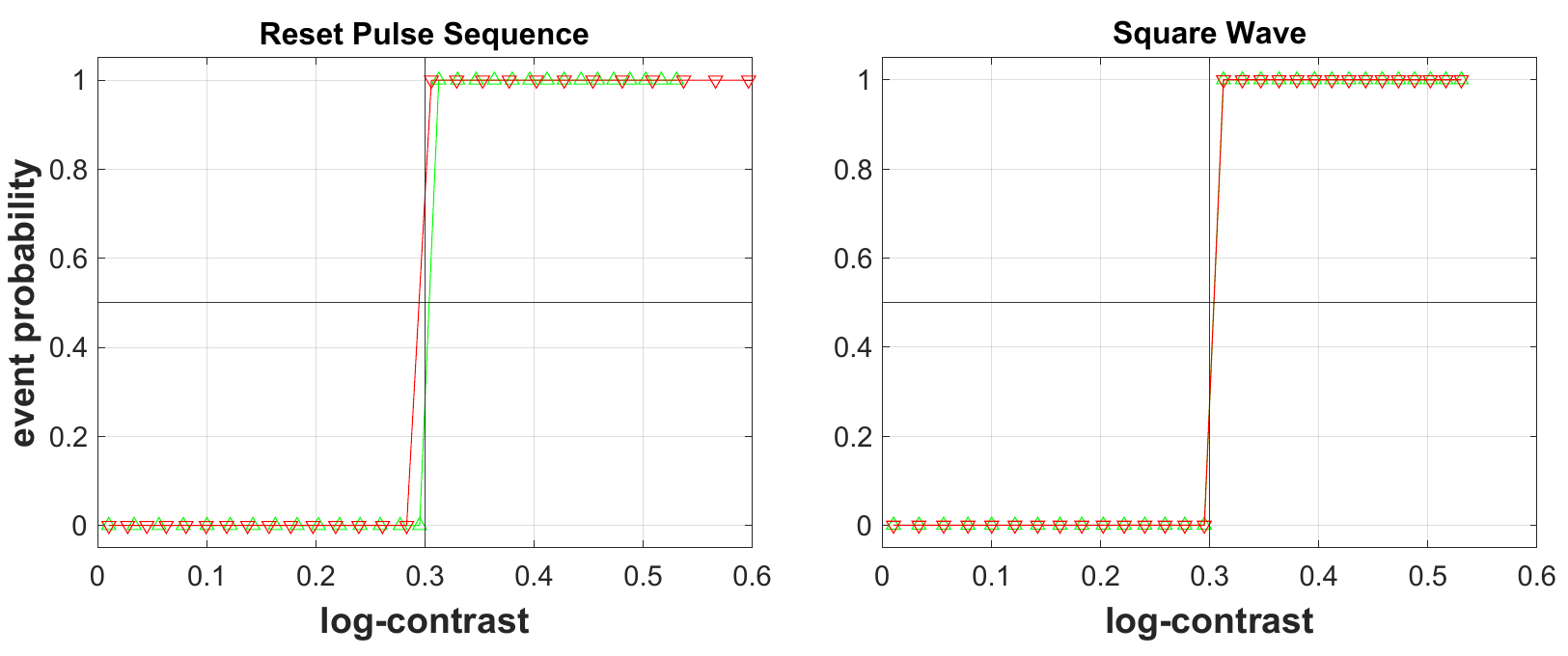}
   \end{center}
   \caption[example] 
   { \label{fig:Ideal} Noise and mismatch free simulation results for $\theta = 0.3$. 'Ideal' conditions result in perfect step response at $\theta$ as predicted.  
 }
   \end{figure}

As shown in \cref{fig:Ideal}, the noise-free case results in a perfect step response as predicted. The measurement error depends only on the resolution of the contrast steps (x-axis), which could be improved by using smaller increments on the signal amplitude between measurements. Both ON (green) and OFF (red) curves cross the 50\% probability threshold within 1\% of the true contrast threshold and there is no discernible difference between the \gls{RP-TP} and \gls{SW} methods.   

\subsection{Adding Noise}
\label{subsec:noise}

\begin{figure} [ht]
   \begin{center}
   \includegraphics[width = \textwidth]{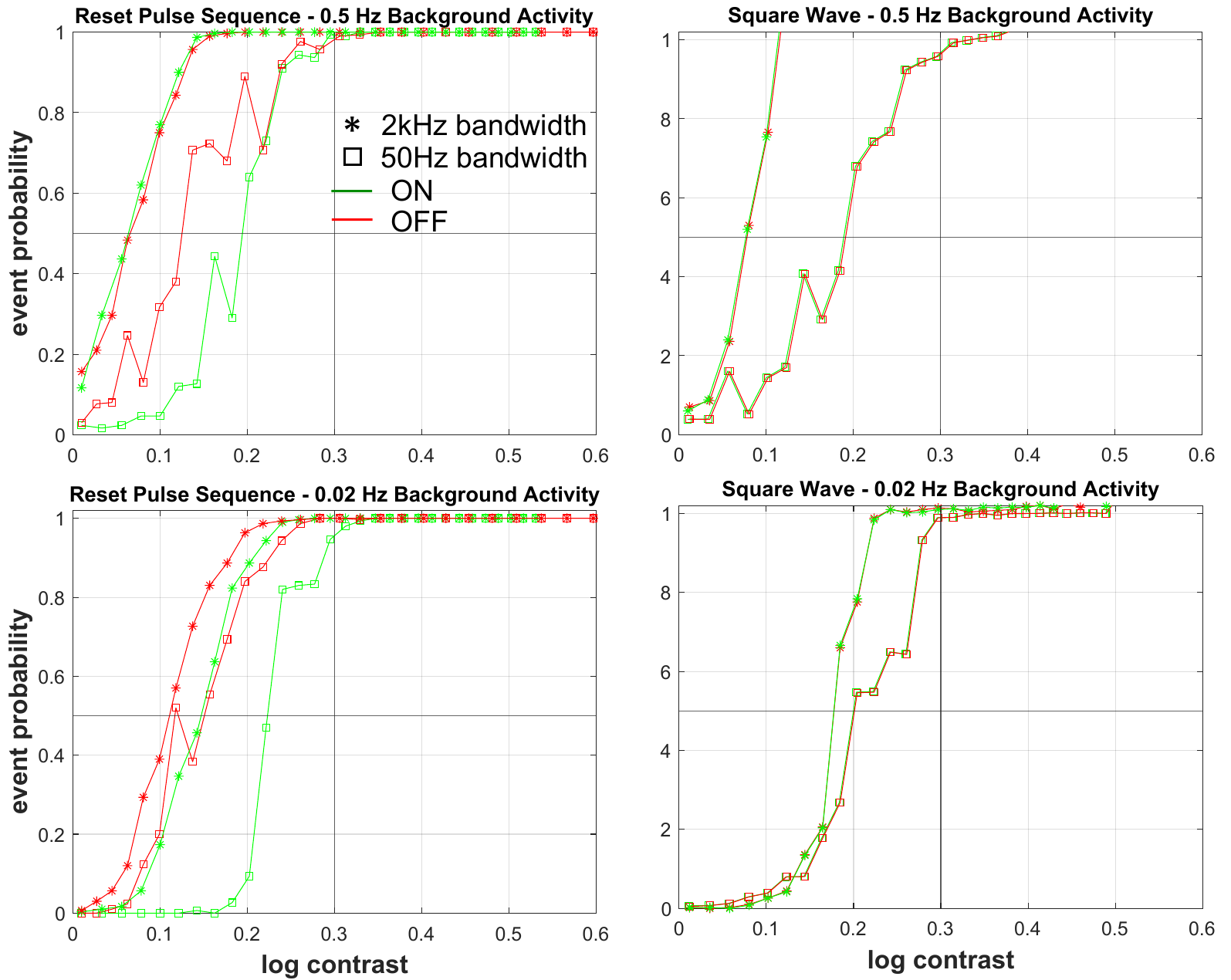}
   \end{center}
   \caption[example] 
   { \label{fig:Noisy} Simulated S-curves with noise for $\theta = 0.3$. Even for low background activity rates, noise shifts the \gls{NCT} measurement (50 \% event probability intercept) to lower contrast than the true threshold level. However, if noise bandwidth is limited (e.g. 50 Hz) by adjusting biases (square datapoints), the 100\% event probability intercept is a better approximation for $\theta$.
 }
   \end{figure}

\cref{fig:Noisy} demonstrates how the S-curve measurement evolves in the presence of noise. Simulated measurements are shown for 0.5 Hz and 0.02 Hz background activity rates for each stimulus waveform. Even with very low background activity rates, \gls{NCT} does not provide a good estimate of $\theta$, and higher noise rates shift \gls{NCT} further to the left. In many applications, high sensitivity (low $\theta$) is desirable. This highlights a \textbf{key shortcoming of the \gls{NCT} metric, as it results in an overly optimistic estimation of the actual threshold levels, and can give the false impression of improved contrast sensitivity in the presence of noise}. Our result suggests that selecting the 50\% event probability intercept is not a good estimate of $\theta$, but still has some utility as a practical metric. On the other hand, \textbf{for an adequately low noise bandwidth (e.g. 50Hz), we see the curves converge to 100\% event probability near the true value of $\theta$}, supporting the interpretation proposed by McReynolds et al. \cite{McReynolds2022-fq}  
Noise bandwidth is optimally controlled by tuning the source-follower bias, although, if accessible in the \gls{EVS} model being tested, it can also be adjusted using the photoreceptor bias\cite{Graca2023-zb}.

To further understand why the Gaussian \gls{CDF} does not accurately characterize a single pixel's event probability curve, we consider a noisy square wave signal. In general, the frequency of noise components is much faster than the frequency of the test stimulus, effectively broadening the signal. As a result, \textbf{noise can only increase the probability that a signal change smaller than $\theta$ results in an event, but will never decrease the probability of an event for a signal change larger than $\theta$}. Response curves generated for \gls{RP-TP} and \gls{SW} stimuli have notable differences; however, there is no clear benefit of one over the other for most bandwidth/background activity combinations.      

\subsection{Mismatch}

\begin{figure} [ht]
   \begin{center}
   \includegraphics[width = \textwidth]{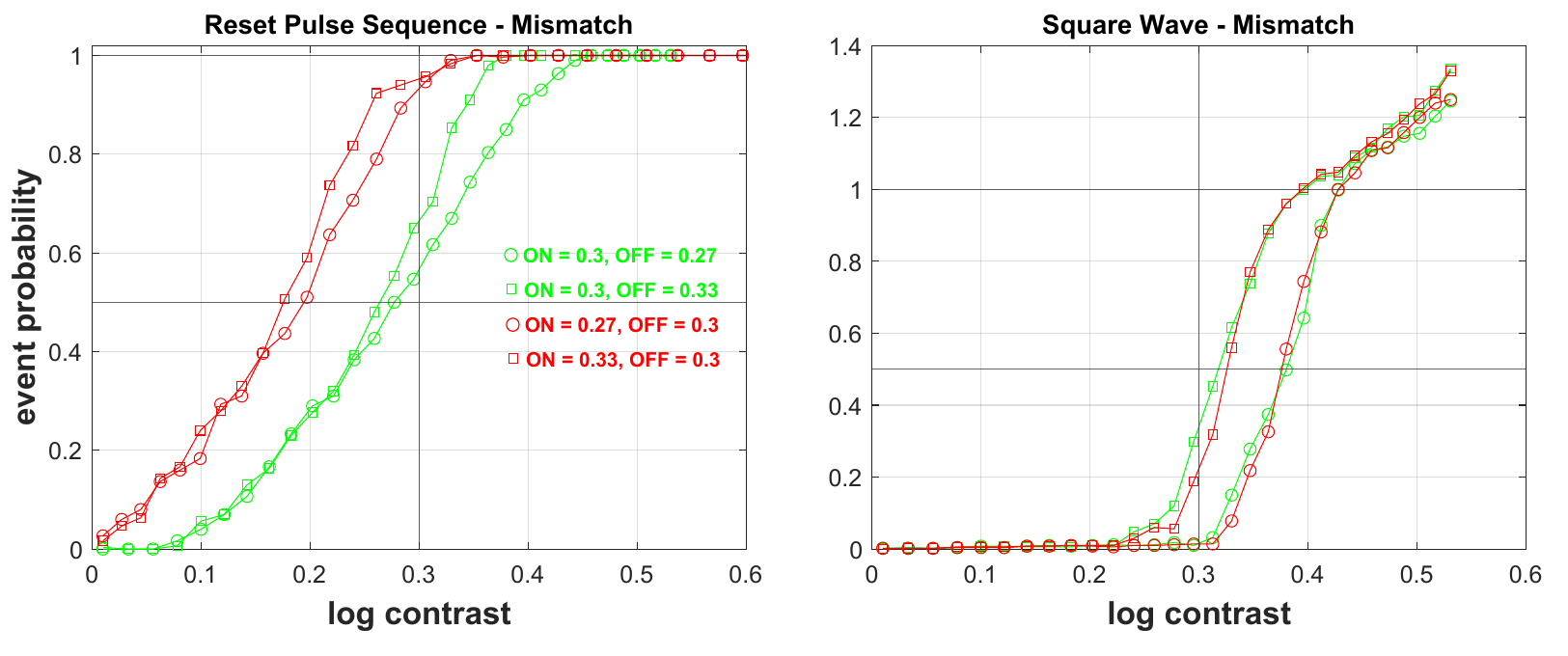}
   \end{center}
   \caption[example] 
   { \label{fig:Mismatch}  Simulation S-curve measurement including mismatch.  When threshold mismatch is applied (i.e. $\theta_\text{on} = \theta_\text{off} + 0.03$) the simulated S-curve shifts to the right and approaches 100\% response probability at a log contrast level higher than the true threshold.
 }
   \end{figure}

In \cref{subsec:noise} we demonstrated that the 100\% intercept of the event probability curve provides an improved estimate of $\theta$, provided that noise bandwidth and background activity rates are properly managed. Next, we explore how measured S-curves are influenced by threshold mismatch.  Because ON and OFF thresholds are determined by independent comparator pathways in the pixel circuit, they are not guaranteed (or likely) to be equal in a particular pixel. Due to device mismatch, typical \gls{EVS} pixels have been shown to exhibit a 1$\sigma$ threshold mismatch of $\approx$ 0.02-0.03 log-contrast. To examine the influence of mismatch, we simulated the 50Hz bandwidth settings from \cref{subsec:noise} and applied a mismatch of 0.03. For both ON and OFF threshold measurement simulations, we examined behavior both when the opposite threshold is higher and lower than the true threshold of the polarity being measured by 1$\sigma$, and the results are shown in \cref{fig:Mismatch}.    

\begin{figure} [ht]
   \begin{center}
   \includegraphics[width = \textwidth]{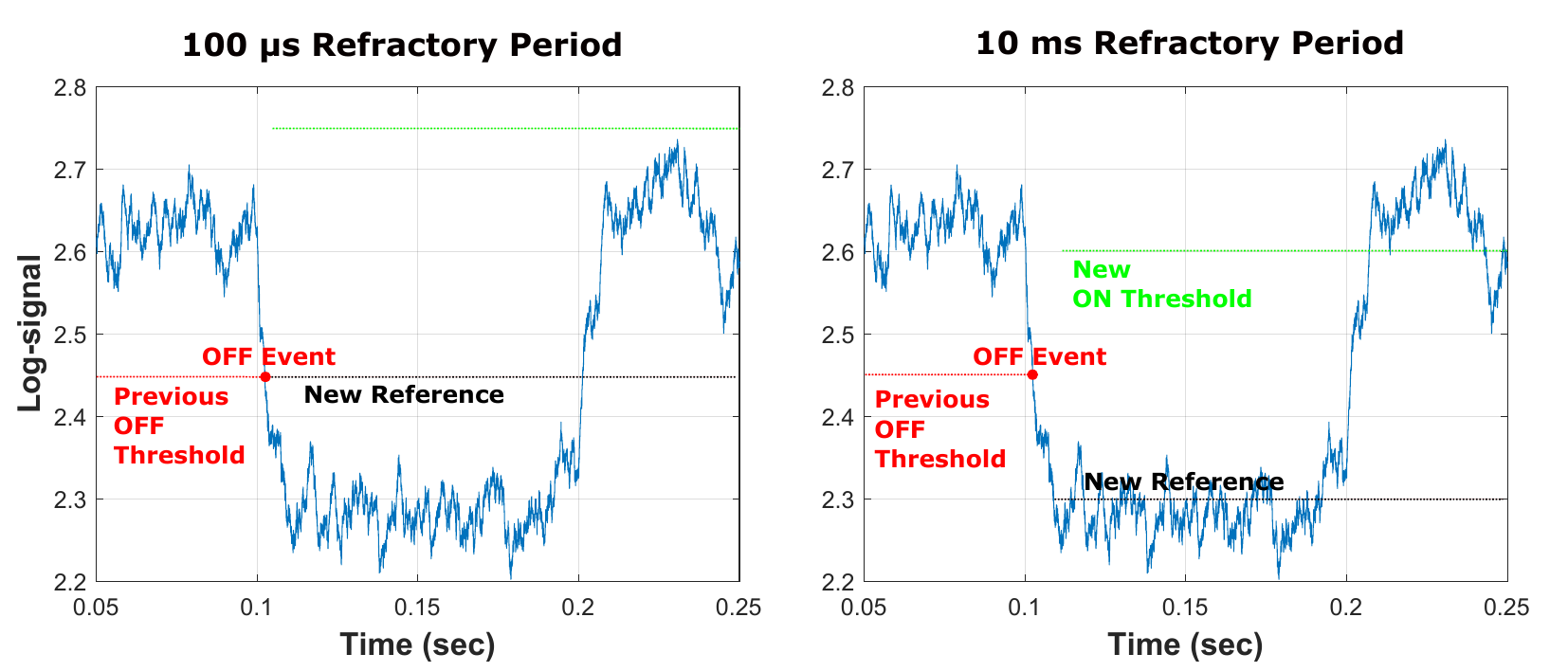}
   \end{center}
   \caption[example] 
   { \label{fig:Mismatch_explanation} One period of simulated square wave. Pixel reset logic and refractory period explain the discrepancy between measured and actual thresholds when threshold mismatch is applied. Increasing the refractory period forces the reset to occur near the base of the stimulus and improves accuracy of the measurement.
 }
   \end{figure}

Observing the 100\% probability intercept, we see the approximation now has shifted to the right of $\theta$, leading to another source of uncertainty in the accuracy of the S-curve measurement. To understand the origin of this behavior, we examine one period of a simulated waveform entering the comparator logic stage.  \cref{fig:Mismatch_explanation} demonstrates how the pixel reset influences whether or not the next event will occur, highlighting another critical consideration for collecting accurate S-curve measurements. On the left, the refractory period is 100 $\mu$s, near the minimum possible setting. To illustrate, we consider an extremely imbalanced case in which $\theta_\text{off} \approx 0.2$ and $\theta_\text{off} \approx 0.3$, and the signal amplitude is slightly larger than the ON threshold. Because the refractory period is nearly instantaneous relative to the bandwidth-limited signal response in the left plot, the new reset level corresponds to the signal level that caused the OFF event rather than the trough of the waveform. The new ON threshold is therefore much higher, meaning a larger signal amplitude is required to consistently generate ON events, leading to the error we observe in the S-curve measurements of \cref{fig:Mismatch}. The plot to the right shows how \textbf{increasing the refractory period (e.g. 10 ms) forces the pixel to wait to reset its reference, resulting in a reference close to the average signal level at the base of the square wave. Under this condition, the event probability provides a more accurate measurement of the signal amplitude relative to the ON threshold}. 

\begin{figure} [ht]
   \begin{center}
   \includegraphics[width = \textwidth]{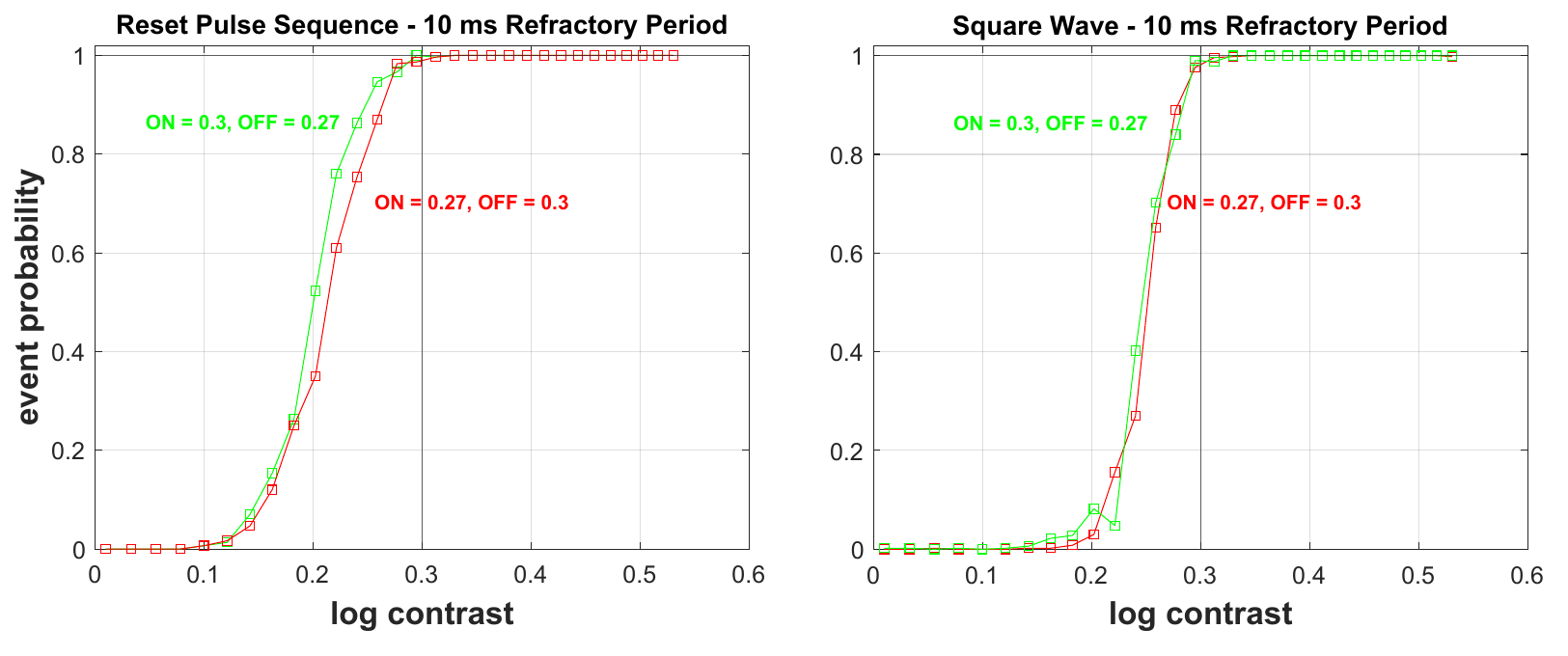}
   \end{center}
   \caption[example] 
   { \label{fig:long_rp} Simulation of S-curve measurement with 10 ms refractory period for $\theta = 0.3$. Noise power resulting in 0.02Hz background activity and 0.03 log contrast mismatch were applied. With both noise and mismatch, increasing the refractory period results in improved accuracy of the S-curve measurement when using the 100\% response to approximate $\theta$.
 }
   \end{figure}

\cref{fig:long_rp} demonstrates how increasing the refractory period improves the S-curve simulation as suggested by analysis of \cref{fig:Mismatch_explanation}. Again, even with mismatch applied, the S-curve approaches 100\% response probability at the larger of $\theta_\text{on}$/ $\theta_\text{off}$. Because mismatch is random, attempting to bias for balanced ON/OFF thresholds leads to uncertainty in which threshold is actually measured for a given pixel. However, since $\theta_\text{on}$ and $\theta_\text{off}$ are able to be tuned individually, \textbf{deliberately setting one lower than the other ensures that the S-curve's 100\% response will correspond to the less sensitive of the two contrast thresholds when using the square wave stimulus}. The \gls{RP-TP} method does not appear to be affected by mismatch as long as the refractory period is sufficiently long. In \cref{subsec:robustness} we explore the merits of this technique and assess the robustness of the prescribed bias settings and measurement procedures.   

\subsection{Robustness of Prescribed Methods}
\label{subsec:robustness}

After adding noise, finite response bandwidth, and threshold imbalance, we have finally developed general biasing techniques to enable accurate S-curve measurements. Bandwidth must be limited (by source follower or photoreceptor bias) to limit noise, a long refractory period ensures the pixel reset level corresponds to the signal baseline prior to each step (pulse), and deliberately applying imbalanced thresholds (square wave method) defines whether the measured 100\% response probability corresponds to $\theta_\text{on}$ or $\theta_\text{off}$. The stimulus light intensity should be high enough for the photocurrent not to limit the pixel bandwidth, and also to minimize noise in the pixel pass-band~\cite{Graca2023-hk}. In \gls{EVS} with light-dependent leak rate\cite{Nozaki2017-px,Graca2023-hk}, however, setting the stimulus light intensity too high results in excessive leakage currents which may significantly affect the S-curve. Finally, in this section we examine how robust these techniques are to both elevated noise rates and large threshold imbalances.

\begin{figure} [ht]
   \begin{center}
   \includegraphics[width = \textwidth]{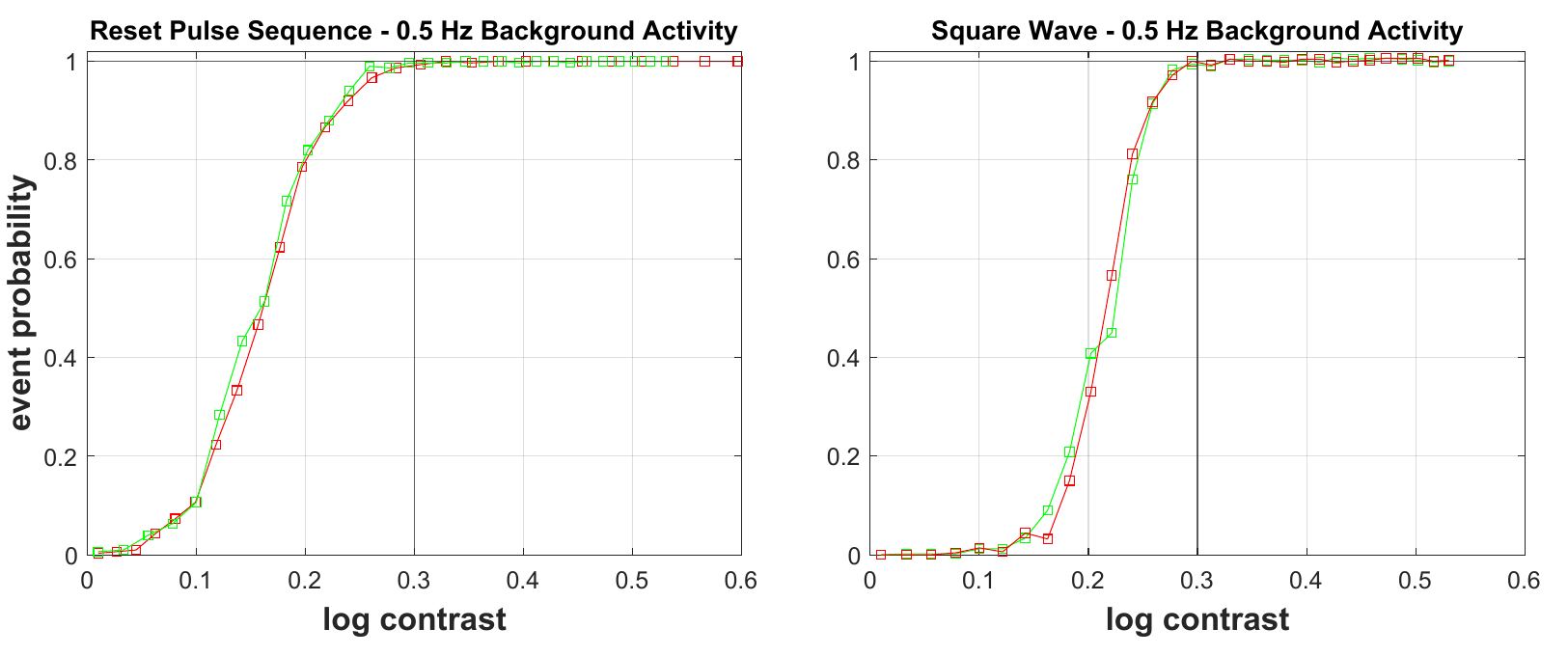}
   \end{center}
   \caption[example] 
   { \label{fig:elevated_noise}  Simulated S-curve with prescribed biases and elevated background activity rate of 0.5 Hz.
 }
   \end{figure}

\cref{fig:elevated_noise} shows that simulated S-curve measurements are relatively robust to increased background activity rates due to higher noise power; however, the square wave stimulus is less affected by the higher noise. Similarly, looking at larger threshold imbalance levels (e.g. the tail ends of the distributions when $\theta_\text{on}$/ $\theta_\text{off}$ are deliberately imbalanced) we observe similar behavior. In both cases, the 100\% probability intercept shifts slightly left for S-curves produced by the \gls{RP-TP} test sequence. Although both methods achieve similar results under near ideal testing conditions (low noise and mismatch), as these non-idealities approach their upper limits, the square wave method outperforms the more complicated \gls{RP-TP} waveform. The \gls{RP-TP} method could be improved/optimized by selecting a more appropriate pre-pulse amplitude, duration, and temporal window for observing the test pulse response. On the other hand, the square wave method produces robust and accurate measurements without any fine-tuning.   

\begin{figure} [ht]
   \begin{center}
   \includegraphics[width = \textwidth]{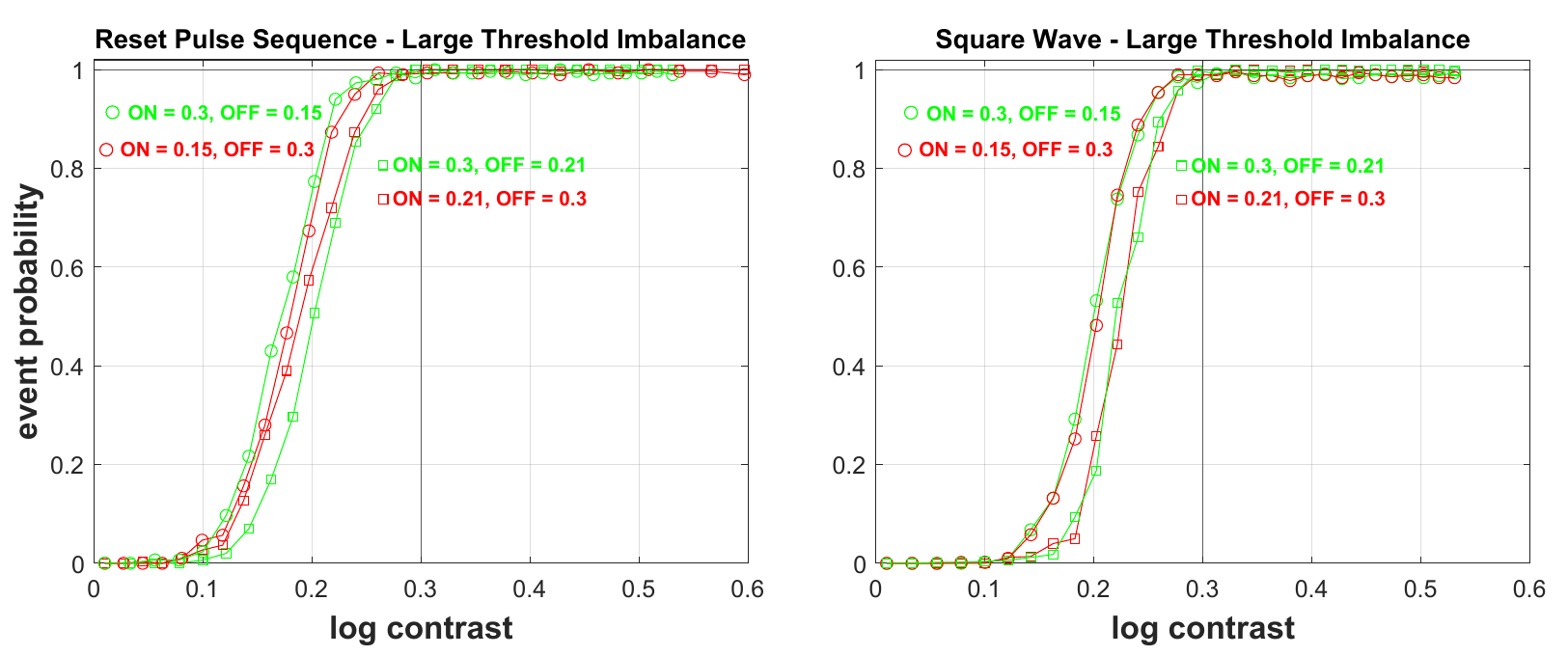}
   \end{center}
   \caption[example] 
   { \label{fig:large_imbal} Simulated S-curve measurement with prescribed biases and larger threshold imbalance.
 }
   \end{figure} 

\section{Camera Measurements}
\label{sec:camera_measurements}

\begin{figure} [ht]
   \begin{center}
   \includegraphics[width = \textwidth]{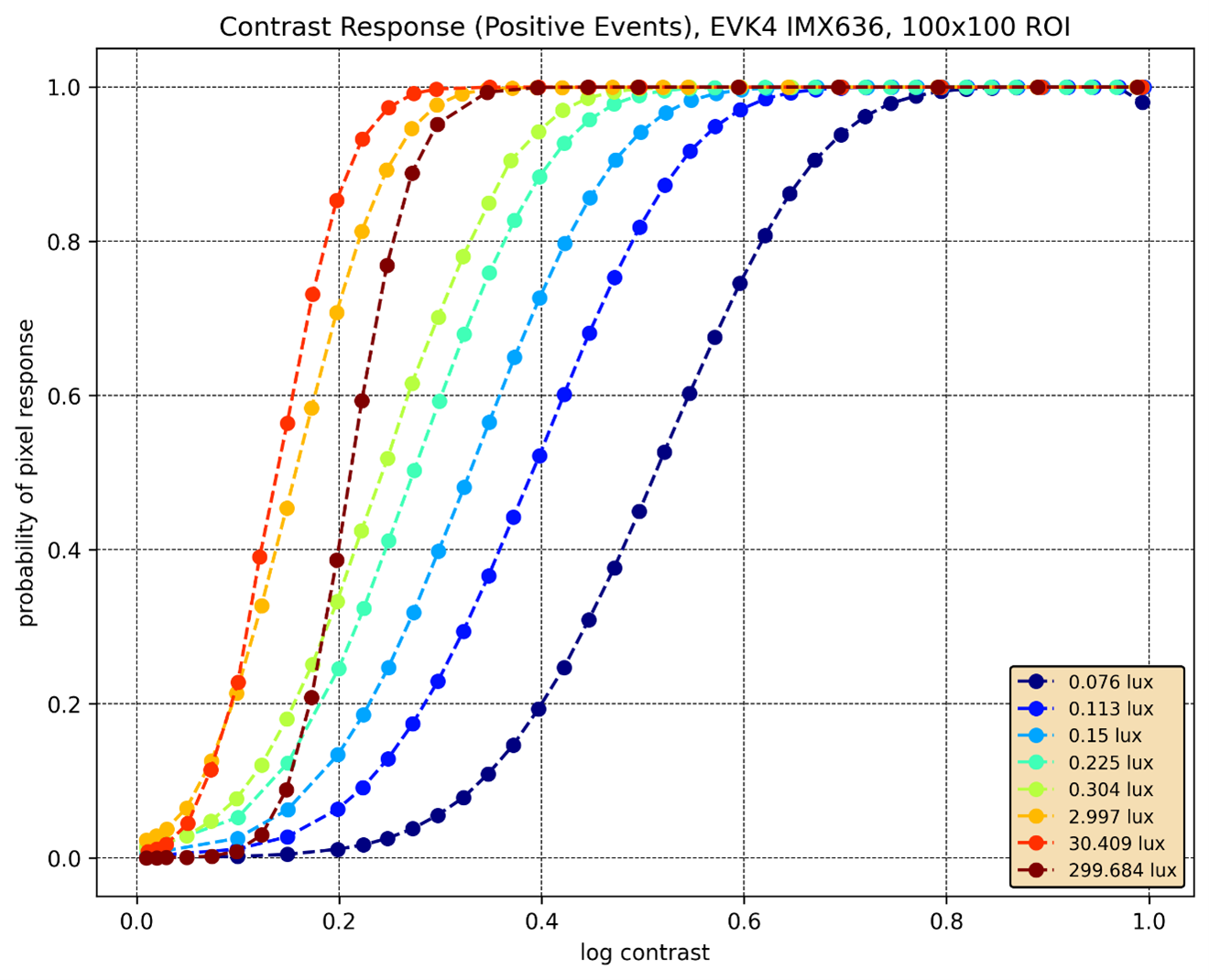}
   \end{center}
   \caption[example] 
   { \label{fig:EVK4_scurves} Average measured S-curves from Sony IMX636 \gls{EVS} using \gls{RP-TP} stimulus procedure with prescribed biasing at different baseline illumination levels for 100$\times$100 pixel region. The 100\% response probability deterministically shifts to higher contrast near the dark current limit, allowing for an estimation of dark current by \cref{eq:on_event}.  
 }
   \end{figure}

To validate our proposed methodology, we generated S-curve measurements for a number of on-chip illumination levels for the Sony/Prophesee EVK4/IMX636, and the result is shown in \cref{fig:EVK4_scurves}. A \gls{ROI} of 100$\times$100 pixels limits readout bandwidth, preventing measurement uncertainty due to readout saturation during each pulse. We used default bias settings other than refractory period bias which was set to -20.  The result was measured for the EVK4 serial number 00050685 attached to a 35 mm f/4 lens. An OD 3.0 ND filter was incorporated to facilitate measurement at the four lowest baseline light levels indicated, mitigating issues with light source flicker and limited radiometric resolution. To measure each S-curve, a train of 100 square pulses was generated using programmatic control of our SpectralLED light source (rise/fall time is about 1 ms). We fine tuned the timing and selected pulse width of about 20 ms. We used the \gls{RP-TP} stimulus method, and used a response window of 40 ms around each test pulse to assess response probability. The curves shown represent the ON event probability.    

\subsection{Inferring Dark Current and Noise}
\label{subsec:dark_current}

Measuring \gls{EVS} thresholds with improved fidelity also has implications for inferring other important performance characteristics necessary to model and predict sensor performance under extreme conditions such as near the dark current limit. Pixel analysis does not predict threshold levels to depend on illumination; however, for illumination levels near the dark current limit, a larger change in photocurrent will be needed to trigger an event. Specifically, an ON event occurs when 
\begin{equation}
\label{eq:on_event}
\ log\left(\frac{I_\text{pho2} + I_\text{dark}}{I_\text{pho1} + I_\text{dark}}  \right) > \theta_{on} \, ,
\end{equation} 
where the signal change is $I_\text{pho2} - I_\text{pho1}$ and $I_\text{dark}$ is the dark current. Therefore, with an accurate measurement of $\theta$, the dark current can be estimated by observing the signal change required to generate an ON event near the dark current limit and rearranging \cref{eq:on_event} to solve for $I_\text{dark}$. 

\begin{figure} [ht]
   \begin{center}
   \includegraphics[width = \textwidth]{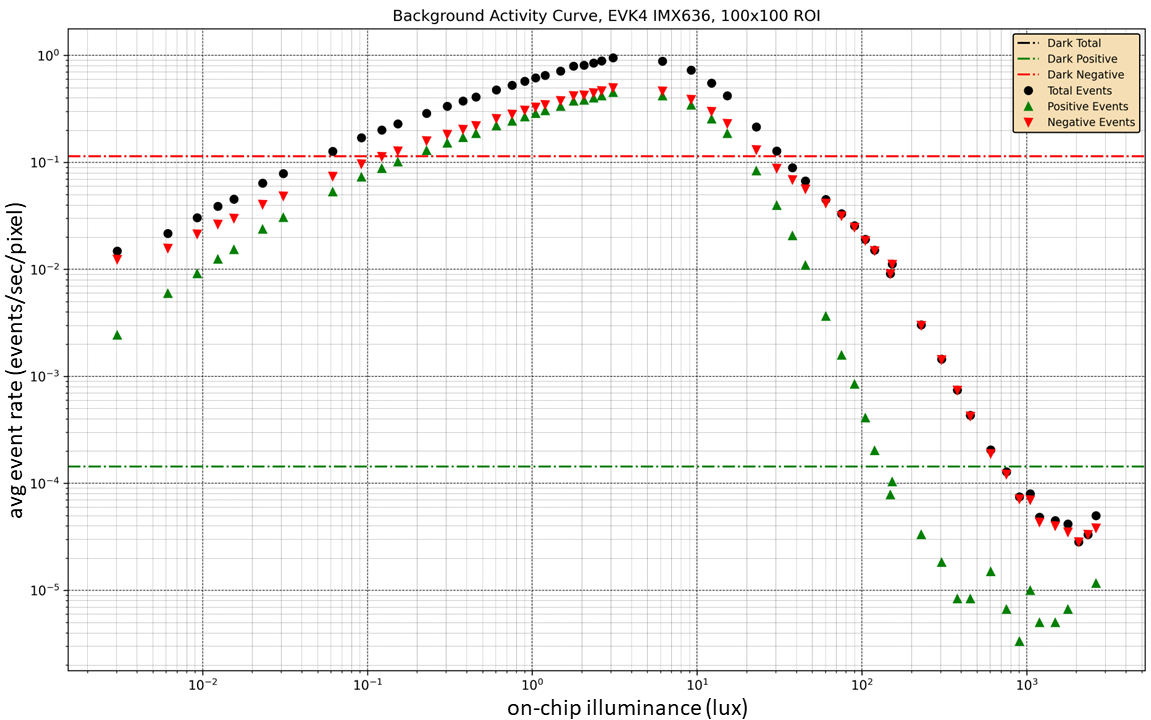}
   \end{center}
   \caption[example] 
   { \label{fig:noise_rates} Sample background activity curve as a function of on-chip illumination for IMX636 with default bias configuration. Dashed lines are steady-state background activity rates for ultra-low illumination levels ($< \SI{1e-5}{}$ lx on-chip.
 }
   \end{figure}

As a result, we propose that a family of S-curves at varied background illumination levels can be used to infer the sensor's dark current. Observing \cref{fig:EVK4_scurves}, the $\theta$ estimate moves progressively to larger log contrast values as the illumination level drops below $\approx$ 300 mlx. At the highest illumination levels (300 lx), when the dark current is much lower than the induced photocurrent, we observe the 100\% probability intercept at $\approx$ 0.35 units log contrast. Observing the noise curve shown in \cref{fig:noise_rates}, we see that noise is elevated at both 3 and 30 lx, which explains why the 100\% probability intercept shifts to the left for those illumination levels. $I_\text{pho1}$, which is the minimum (baseline) photocurrent for the ON threshold measurement, is calculated from baseline on-chip illuminance ($E_\text{v1}$) as 
\begin{equation}
\label{eq:lux2Iph}
I_\text{pho1} = \SI{1.12e16}{}E_\text{v} \eta q p^2 \, ,
\end{equation} 
where $\eta$ is average spectral quantum efficiency, $p$ is pixel pitch, $\SI{1.12e16}{}$ the scalar conversion between lumens and photon flux, and q the electron charge. We used specs from a similar \gls{CMOS} image sensor by Sony to estimate quantum efficiency. $I_\text{pho2}$ is a scalar multiple of $I_\text{pho1}$ representing the linear change in illumination. For example, observing the 76 mlx curve in \cref{fig:EVK4_scurves}, a log contrast of 0.8 is required to achieve $\approx$ 100\% response probability; therefore, we infer $\log\left(\frac{I_\text{pho2}}{I_\text{pho1}}\right) = \log\left(\frac{E_\text{v2}}{E_\text{v1}}\right) = 0.8$, or rather, $I_\text{pho2} = I_\text{pho1} e^{0.8}$. Solving \cref{eq:on_event} for $I_\text{dark}$ at the five lowest illumination levels (76 - 304 mlx) we find close agreement across all curves with $I_\text{dark} = 5.55 \pm 0.28$ fA. 

Though dark current is not reported for the IMX636, \gls{LLCO} is defined as the minimum light level required for 50\% of pixels to respond to a linear, 100\% contrast step. For an estimated $\theta_\text{on} = 0.25$ units log contrast, \gls{LLCO} is reported as 80 mlx. Given the definition of \gls{LLCO}, we make use of \cref{eq:on_event} by setting $I_\text{pho2} = 2 I_\text{pho1}$ to check our independent $I_\text{dark}$ estimate and find $5.62$ fA, increasing confidence in the accuracy of our proposed method.

One caveat of this estimation is the possibility that low pixel bandwidth near the dark current limit attenuates the signal amplitude, leading to an overestimation of the dark current. For example, if the pixel's front-end, low-pass corner frequency is $\approx$2Hz near the dark current limit, the signal pulse will only reach 40\% of its peak value within 40 ms of each test pulse. The temporal response near the dark current has not been well characterized for the IMX636, but reported latency at 5 lx (1 ms) suggests a corner frequency that monotonically depends on illumination could be on the order of single Hz near the dark current limit. Further tests are required to characterize and isolate the effects of pixel bandwidth to improve the dark current estimation.   

Another important benefit of correct interpretation and improved methodology of the S-curve measurement is the ability to devise realistic simulations including physical noise parameters. \cref{fig:noise_rates} includes background activity rates as a function of on-chip illuminance. Now that we have confirmed that the threshold itself does not vary with illumination, we can correctly infer that the change in background activity is entirely dependent on the noise profile at a particular illumination level. A brute force approach would be to simulate background activity at various levels of additive noise power to map noise power to illumination, but this ignores the key ingredient of noise bandwidth. A more thorough analysis should also account for the illumination dependent pixel bandwidth; however, that level of analysis is beyond the scope of this paper. Regardless, accurate threshold measurements are the crucial first step for improved modeling of pixel behavior.

\section{Conclusion}
\label{sec:conclusion}

Contrast threshold, $\theta$, is the most important parameter for defining \gls{EVS} response, and the pixel response probability curve, or S-curve is an key method used to estimate $\theta$. In this work, we explored two methods to obtain S-curves and analyzed the resulting curves. Using a pixel simulation, we showed that the commonly accepted interpretation of the S-curve does not actually approximate $\theta$ when subject to real world physical constraints such as noise and mismatch. Our simulation results validate a prior suggestion that the true threshold is better approximated by observing where the S-curve approaches 100\% response probability \cite{McReynolds2022-fq}. Additionally, we show that the approximation is subject to error unless the sensor is carefully biased to manage noise bandwidth and threshold mismatch. As a result we prescribe a bias configuration with low-bandwidth, long refractory period, and deliberately mismatched thresholds where the higher setting corresponds to the threshold being measured. We also compared methods using a simple square wave stimulus and a more complicated \gls{RP-TP} waveform. Our results show that both are viable; however, the simple square wave demonstrated more robustness to noise and mismatch. Finally, we demonstrated that accurate measurements along with our improved interpretation of the S-curve makes it possible to estimate dark current by observing how the response curves change near the dark current limit. Our results are important for researchers exploring new applications of \gls{EVS} near the edge of their performance limits, and useful for developing high fidelity \gls{EVS} simulations for both new and existing commercial sensor models.            

\appendix    

\acknowledgments 
 
Rui Graca is funded by the University of Zurich, Swiss National Science Foundation project SCIDVS (Grant No. 185069).  

\bibliography{paperpile} 
\bibliographystyle{spiebib} 

\end{document}